\documentclass{article}

\usepackage[preprint, nonatbib]{neurips_2026}
\usepackage[numbers]{natbib}
\newif\ifsubmission
\submissiontrue

\usepackage{amsmath,amsfonts,bm}

\def\eqref#1{equation~\ref{#1}}

\def\1{\bm{1}}

\DeclareMathAlphabet{\mathsfit}{\encodingdefault}{\sfdefault}{m}{sl}
\SetMathAlphabet{\mathsfit}{bold}{\encodingdefault}{\sfdefault}{bx}{n}

\usepackage{hyperref}
\usepackage{url}
\usepackage{subcaption}
\usepackage{wrapfig}
\usepackage{microtype}
\usepackage{adjustbox}
\usepackage{balance}
\usepackage{booktabs}
\usepackage{graphicx}
\usepackage{subcaption}
\usepackage{siunitx}
\usepackage{tcolorbox}
\tcbuselibrary{breakable}

\usepackage{amsmath,amssymb}
\usepackage{algorithm,algpseudocode}
\usepackage{xspace}
\usepackage{listings}
\definecolor{mygray}{HTML}{e3e6e8}
\definecolor{darkgreen}{RGB}{0,120,0}

\newcommand{\graybox}[1]{\begin{tcolorbox}[breakable,width=\linewidth,boxrule=1pt,top=0pt,bottom=0pt, left=1pt,right=1pt]#1\end{tcolorbox}}
\newcommand{\code}[1]{{\setlength\fboxsep{1pt}\colorbox{mygray}{\texttt{\lstinline|#1|}}}}

\ifsubmission
  \newcommand{\todo}[1]{\ignorespaces#1\unskip}
  \newcommand{\new}[1]{#1}
  
  \newcommand{\todel}[1]{}
  \newcommand{\toadd}[1]{\ignorespaces#1\unskip}
  \newcommand{\tochk}[1]{}
  \newcommand{\bo}[1]{}
  \newcommand{\ruishi}[1]{}
  \newcommand{\prateek}[1]{}
  \newcommand{\umang}[1]{}
  \newcommand{\brandon}[1]{}
\else
  \newcommand{\todo}[1]{\textcolor{teal}{\ignorespaces#1\unskip}}
  \newcommand{\new}[1]{#1}
  
  \newcommand{\todel}[1]{\textcolor{red}{#1}}
  \newcommand{\toadd}[1]{\textcolor{darkgreen}{#1}}
  \newcommand{\tochk}[1]{\textcolor{gray}{TOCHK: #1}}
  \newcommand{\bo}[1]{\textcolor{brown}{Bo: #1}}
  \newcommand{\ruishi}[1]{\textcolor{blue}{Ruishi: #1}}
  \newcommand{\prateek}[1]{\textcolor{red}{Prateek: #1}}
  \newcommand{\umang}[1]{\textcolor{cyan}{Umang: #1}}
  \newcommand{\brandon}[1]{\textcolor{orange}{Brandon: #1}}
\fi

\algrenewcommand\algorithmicrequire{\textbf{Input:}}
\algrenewcommand\algorithmicensure{\textbf{Output:}}
\algnewcommand\algorithmicparams{\textbf{Parameters:}}
\algnewcommand\Params{\Statex\hspace*{-\algorithmicindent}\algorithmicparams\ }

\newcommand{\our}{\text{ACToR}\xspace}
\newcommand{\ournf}{\text{ACToR-NoFuzz}\xspace}
\newcommand{\tagent}{\mathcal{T}\xspace}
\newcommand{\dagent}{\mathcal{D}\xspace}

\newcommand{\setourSWEGPT}{\our@m-SWE GPT-5-mini\xspace}
\newcommand{\setourSWECLAUDE}{\our@m-SWE Claude4\xspace}

\tcbuselibrary{skins,breakable} 
\lstdefinelanguage{Markdown}{
  morecomment=[l]{\#},
  morestring=[b]",
}

\tcbset{
  promptbox/.style={
    colback=gray!5,
    colframe=black,
    boxrule=0.5pt,
    arc=1.5mm,
    fonttitle=\scriptsize\bfseries,   
    width=\linewidth,
    left=3mm,
    right=3mm,
    top=1mm,
    bottom=1mm
  }
}

\usepackage[utf8]{inputenc} 
\usepackage[T1]{fontenc}    
\usepackage{hyperref}       
\usepackage{url}            
\usepackage{booktabs}       
\usepackage{amsfonts}       
\usepackage{nicefrac}       
\usepackage{microtype}      
\usepackage{xcolor}         

\title{Adversarial Agent Collaboration for Correctness Improvements of C to Safe Rust Translation}

\author{%
  Tianyu Li\textsuperscript{1} \quad
  Ruishi Li\textsuperscript{1} \quad
  Bo Wang\textsuperscript{1} \\[0.35em]
  \bfseries
  Brandon Paulsen\textsuperscript{2} \quad
  Umang Mathur\textsuperscript{1} \quad
  Prateek Saxena\textsuperscript{1} \\[0.7em]
  \normalfont
  \textsuperscript{1}National University of Singapore, Singapore \qquad
  \textsuperscript{2}Amazon Web Services, USA \\[0.5em]
  \normalfont\texttt{\{tianyuli, liruishi, bo\_wang\}@u.nus.edu} \\[0.15em]
  \texttt{bpaulse@amazon.com} \\[0.15em]
  \texttt{\{umathur, prateeks\}@comp.nus.edu.sg}%
}

\begin{document}

\definecolor{darkgreen}{RGB}{0, 120, 0}  
\newcommand{\fix}[1]{#1}

\maketitle


\begin{abstract}
Translating C to memory-safe languages, like Rust, prevents critical memory safety vulnerabilities that are prevalent in legacy C software.
\toadd{Even with recent LLM-based and tool-augmented translators, the resulting Rust code frequently diverges from the C source on inputs absent from the test suite used during translation; this correctness gap on \emph{unseen} inputs remains a dominant obstacle to reliable, automatic C-to-Rust translation.}
\toadd{In this work, we present \our (\textbf{A}dversarial \textbf{C} \textbf{To} \textbf{R}ust), a simple LLM-agent loop that closes this gap by adversarially searching for inputs on which the translation diverges from the C source, and using them to drive subsequent refinements.}
\toadd{Inspired by GANs, \our pits a translator agent against a discriminator agent that collaborate to iteratively refine the Rust translation.}
On each iteration, the translator agent synthesizes and refines a Rust translation to pass an existing suite of tests, and then the discriminator agent finds new failing tests \toadd{ by constructing and refining a differential fuzzer over the C and Rust binaries}.
\toadd{Across 63 real-world command-line C utilities, with an average size of 473 lines of code and the longest reaching thousands of lines in size, \our achieves over 90\% test pass rate with zero human intervention.}
\toadd{The improvement holds across seven agent--LLM configurations on our micro-benchmark, indicating that the loop is largely independent of the choice of underlying translator and LLM.}
\toadd{Compared to a non-adversarial, coverage-driven test-generation baseline, \our improves correctness by up to 36.7\%.}
\toadd{
When applied on top of one recent translator, C2SaferRust, \our further improves the validation pass rate by 16.6\%.
}
\end{abstract}



\section{Introduction}

Memory safety vulnerabilities in C/C++ code constitute a 
large fraction of security vulnerabilities each year.
Microsoft reported 70\% of their CVEs are due to memory safety issues~\citep{microsoftproactive,microsoftsaferlang}.  
In response, there has been an increasing demand to translate
memory-unsafe legacy code to modern memory-safe languages like Rust from industry and governments~\citep{darpa2025tractor},
thereby ensuring full memory safety by design.
Unfortunately, given the scale of legacy code that permeates our software systems today,
manual translation of (millions of lines of) 
existing C code is a formidable and practically infeasible task.
A reasonable translation from C to Rust must use safe and idiomatic Rust programming abstractions, 
which are checked mostly at compile time and avoid runtime performance overhead, 
unlike memory safety checking directly in C~\citep{cets2010,softbound,cetssb2024,asan}. 



\toadd{Existing approaches fall into two categories. Rule-based translators (e.g., c2rust~\citep{c2rust} and corrode~\citep{corrode}) produce largely unsafe Rust; converting this unsafe output into safe, idiomatic Rust remains a major open challenge, with substantial follow-up work on type inference~\citep{emre2023aliasing, zhang2023ownership, xu2025typemigration, tymcrat}, idiom-specific refactoring~\citep{taggedunion, outputparam, fileapi}, and syntactic rewriting~\citep{ling2022rust}. However, these techniques typically address only parts of the conversion, and producing fully safe Rust output often still requires manual effort. LLM-based approaches---including both single-shot translation and agentic setups that iterate on compiler errors and test failures during translation---produce safer, more idiomatic Rust, but they share a fundamental correctness gap: the translated code often diverges from the C source on inputs not exercised by the tests available during translation~\citep{ruishi2024ndss, pan2024lost}, and several systems report needing manual intervention or specialized analyses~\citep{syzygy, cai2025rustmap, nitin2025c2saferrust, zhou2025llm, farrukh2025safetrans}.}

\toadd{This is fundamentally a generalization problem: with only a limited test set available during translation, the translator cannot easily tell whether its output matches the C source's behavior on unseen inputs. Augmenting the agent with standard line-coverage tools~\citep{gcov} does not close the gap, as we show in our evaluation: such tests, while achieving high line coverage, are limited in revealing the semantic divergences that drive translation toward correctness. Closing this gap requires techniques that adaptively target divergences between the candidate Rust translation and the C reference program.}

\begin{figure*}[t]
\centering
\includegraphics[width=0.9\textwidth,trim=0 9.8cm 9.8cm 0.3cm]{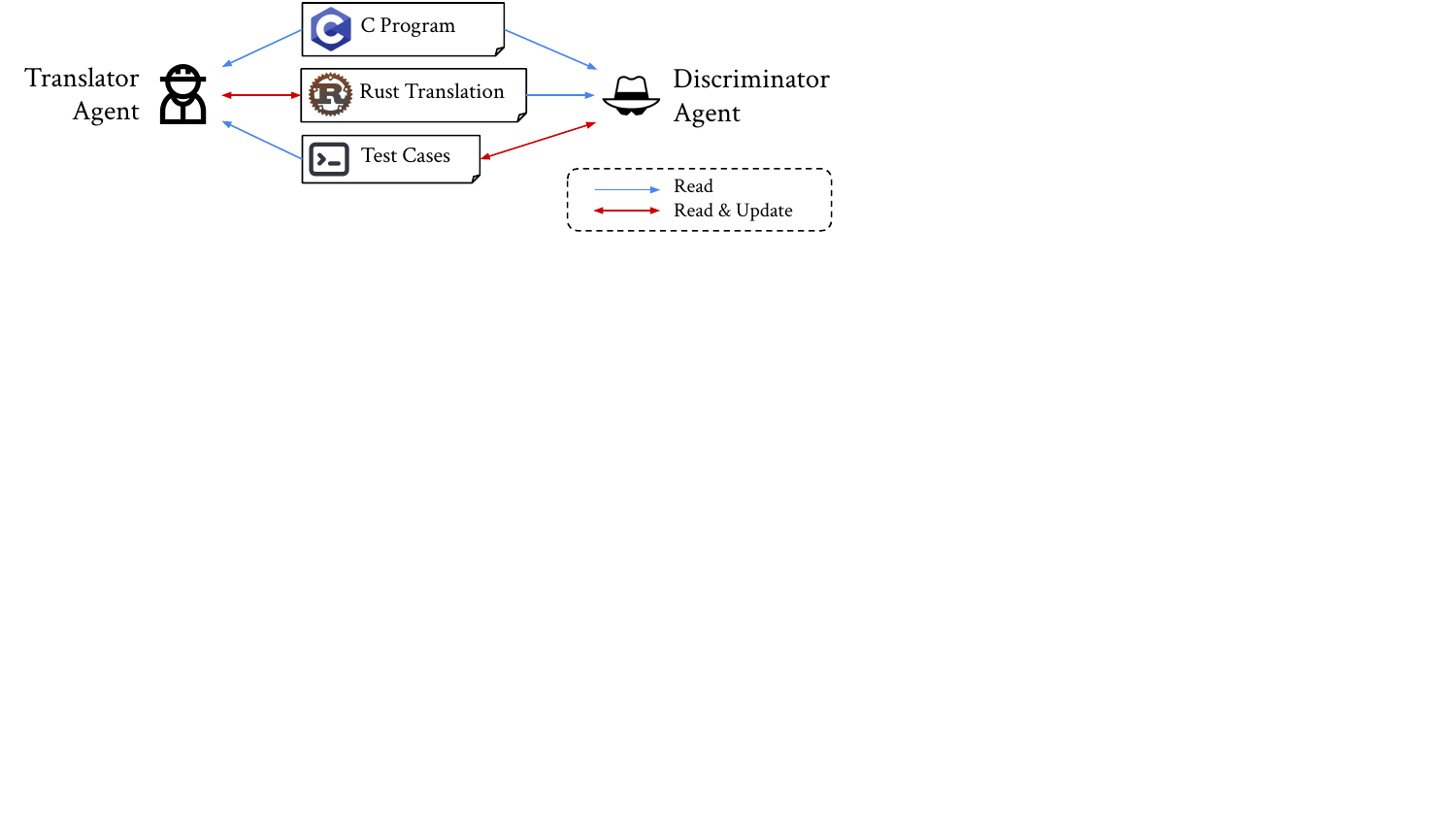}  
\caption{
High-level overview of \our. The Translator and the Discriminator agents update the translation and the tests in turn to iteratively improve the correctness of the translated program.
}
\label{fig:overview}
\end{figure*}

In this work, \toadd{we introduce an adversarial loop that wraps a translator agent and overcomes the limited generalization observed in non-adversarial agent frameworks.} 
\toadd{Inspired by the generator--discriminator paradigm of GANs~\citep{GANGoodfellow2020}, we employ two collaborating agents: a \emph{translator}, which proposes candidate Rust translations, and a \emph{discriminator}, which actively searches for evidence (in the form of counterexamples) that the translation diverges from the C source.}
Whenever the discriminator identifies such adversarial inputs, 
these are fed back to the translator, which uses them to 
refine subsequent translations. 
\toadd{Over iterations, the translator's output not only passes the original tests but also withstands the discriminator's adversarial probing.}
We refer to this adversarial translator–discriminator \toadd{loop} as {\bf \our},
for Adversarial C To Rust, as demonstrated in Figure~\ref{fig:overview}.

We evaluate \our on two sets of benchmarks: A micro-benchmark set of $6$ C programs considered in recent work, and a macro-benchmark set of $57$ C programs from BSDCoreUtils. \new{\toadd{These $63$ real-world C programs average 473 lines of code, longest in the thousands, totaling around 30k.}
\our automatically produces Rust translations for all benchmarks that pass more than 90\% of tests on average with zero human intervention during translation.
\toadd{The improvement holds across seven agent--LLM configurations on our micro-benchmark, indicating that the loop is largely independent of the underlying translator and LLM.}}
\toadd{Applied on top of C2SaferRust's translations on \todo{7} standalone programs from its benchmark, \our raises pass rate by approximately \todo{16.6\% (from 76.3\% to 92.9\%)}, showing \our can complement existing translators to improve correctness.}
%
%

In addition, we show that \our's adversarial design choice improves pass rates by up to \toadd{36.7}\% compared with \toadd{a non-adversarial, coverage-driven test-generation baseline}.
\toadd{Against a strong baseline (Claude Opus 4.6 with no adversarial loop), \our with a weaker LLM (Claude-Sonnet-4.5) reaches 98.2\% pass rate on the micro-benchmark vs.\ the strong baseline's 94.7\%, and on the macro-benchmark wins on 32/57 programs (vs.\ 3/57; 22 non-discriminating ties). This suggests the adversarial loop can compensate for differences in underlying LLM capability.}
\toadd{The adversarial loop is also not tied to proprietary models: applied with the open-weights GLM~4.7 (via Mini-SWE-Agent), \our reaches 95.3\% pass rate on the micro-benchmark, comparable to the strong baseline's 94.7\%.}
\toadd{Our code, benchmarks, and resulting translations are provided as supplementary material and will be released publicly upon publication.}

%



\section{Related Work}
\label{sec:background}

Automated translation of C to Rust is a relatively recent approach towards achieving full memory safety~\citep{emre21}. The desired goal of such translation is to generate safe\footnote{The use of \texttt{unsafe} keyword in Rust allows mixing unchecked and possibly memory unsafe code in otherwise safe Rust} Rust code corresponding to C code, i.e., Rust code that uses only idiomatic Rust primitives that the compiler can statically verify or trust to be memory safe. \toadd{While several prior works have made progress along distinct objectives (e.g., type inference, scalability via decomposition), achieving full correctness of the translated Rust code remains a major challenge~\citep{ruishi2024ndss}.}

\subsection{C to Rust Translation}
\label{sec:related-translation}

Existing works on C-to-Rust translation can primarily be
categorized as rule-based or LLM-based techniques. 
Rule-based techniques such as c2rust~\citep{c2rust} and corrode~\citep{corrode} 
opt for mechanical, line by line, rewriting of the C code, 
often resorting to raw pointers and unsafe Rust. 
Later efforts explored ways to infer Rust types from C code, either by compiler 
feedback~\citep{emre2023aliasing,tymcrat}, SMT solving~\citep{zhang2023ownership},
 data-flow graphs~\citep{xu2025typemigration}, or 
syntactic rewriting~\citep{ling2022rust}, and 
specialized analyses targeted idioms like tagged unions~\citep{taggedunion}, 
output parameters~\citep{outputparam}, and file APIs~\citep{fileapi}. 
Beyond general C, Low* compilation to Rust~\citep{fromherz2024compile} 
demonstrates scalable translation in restricted settings, 
though such techniques remain infeasible for general-purpose 
C due to expressiveness and unmodeled APIs.

LLM-based approaches use an LLM to propose a translation~\citep{lachaux2020unsupervised}. Recent work augments models with different kinds of tools. These include test generation tools~\citep{eniser2024towards,yang2024vert}, dynamic pointer analyses~\citep{syzygy}, or multi-modal prompting~\citep{nitin2024spectra}. Function-level decomposition has been explored to improve scalability~\citep{smartc2rust,cai2025rustmap,ou2025enhancing}, while agentic workflows combine LLMs with static analysis~\citep{zhou2025llm,farrukh2025safetrans}. Neural methods yield more idiomatic code, yet often fall short of preserving exact input/output behavior. \toadd{Moreover, behavior often diverges from the C source on inputs outside the test suite used during translation.} Studies echo these concerns: decomposition introduces inconsistencies~\citep{ruishi2024ndss}, and LLMs remain prone to subtle errors~\citep{pan2024lost}.

\toadd{Existing C-to-Rust systems make distinct design tradeoffs. Some prioritize scalability through decomposition or static analysis~\citep{nitin2025c2saferrust, cai2025rustmap}; others target fully safe Rust output but do not aim for compilable end-to-end translations~\citep{tymcrat}; yet others trade automation for correctness via dynamic profiling and human-in-the-loop refinement~\citep{syzygy}. Table~\ref{tab:related} positions these systems along several dimensions. \textbf{To our knowledge, \our is the only system in this space that targets both fully safe Rust and correctness on inputs unseen during translation without human intervention.} The adversarial loop is also complementary to these systems: any of the LLM-based translators in the table can in principle serve as the underlying translator that \our's discriminator augments.}

\begin{table}[h]
\centering
\caption{\our positioned against representative LLM-based prior C-to-Rust systems. ``Translator'' distinguishes single-shot LLM calls (LLM) from interactive coding agents (Agent). ``Correctness'' indicates whether reported pass rates are measured on tests held back from the translator (\emph{hidden tests}) or shared with it (\emph{given tests}). ``--'' marks dimensions that do not apply or are not reported.}
\label{tab:related}
\resizebox{\textwidth}{!}{%
\begin{tabular}{lcccccr}
\toprule
System & Analysis & Translator & No Human & Safe Rust & Correctness & Eval Size \\
\midrule
\our (this paper)              & --          & Agent & \checkmark & \checkmark & Hidden tests, $\sim$95\%  & $\sim$30k LoC \\
Tymcrat~\citep{tymcrat}        & Lightweight & LLM   & \checkmark & \checkmark & -- (signatures)           & $\sim$800k LoC \\
C2SaferRust~\citep{nitin2025c2saferrust} & Static & LLM & \checkmark & --      & Given tests, partial      & $\sim$270k LoC \\
Syzygy~\citep{syzygy}          & Dynamic     & LLM   & --         & \checkmark & Given tests, partial      & $\sim$3k LoC \\
EvoC2Rust~\citep{evoc2rust}    & Lightweight & Agent & --         & Mostly     & Given tests, partial      & $\sim$8k LoC \\
SmartC2Rust~\citep{smartc2rust} & --         & LLM   & \checkmark & Mostly     & Given tests, full         & $\sim$40k LoC \\
IRENE~\citep{irene}            & Static      & LLM   & \checkmark & Mostly     & Given tests, partial      & Small \\
\bottomrule
\end{tabular}%
}
\end{table}

\subsection{Collaborative LLMs \& Agents}
Many works propose multi-LLM and multi-agent systems, where the LLMs or agents collaborate to produce a final answer. Many works on code generation propose specific roles, such as coder, tester, or analyst, which collaborate to write code~\citep{mcaleese2024llmcritics, chencodet, dong2024selfcollab, huang2023agentcoder, qian2024chatdev, islam2024mapcoder}. Collaborative approaches have been used outside of code generation as well. Many works propose self-reflection techniques~\citep{shinn2023reflexion, madaan2023selfrefine, dhuliawala2023cove}, sometimes aided by tools~\citep{pan2024lost, goucritic, dongselfplay}, where the LLM first proposes an answer, then reflects on its answer to improve accuracy. Similar works propose multi-agent debate architectures~\citep{chan2024chateval, du2023multiagentdebate, liang2023multiagentdebate2, hong2024metagpt,wu2024autogen, liu2024dynamic} where multiple answers are proposed, then the LLM(s) debate the correct answer. Several works propose frameworks for agent collaboration~\citep{hong2024metagpt,wu2024autogen,liu2024dynamic}. 

The key difference between ours and prior work is in the high-level principle: To reliably improve the translated code, we leverage the C program as the {\em oracle} in the construction of a discriminator, which automatically identifies correctness gaps and provide iterative feedback to the translator.

\section{\our Design}
\label{sec:design}

\newcommand{\iseqTxt}{\texttt{IsEq}}
\newcommand{\iseq}[2]{\iseqTxt(#1, #2)}

\toadd{Given a C program and a seed test set, \our produces a safe Rust translation that generalizes beyond the seeds. This section formalizes the translation problem, identifies the generalization gap, and introduces the adversarial loop that narrows it.}

\subsection{\fix{Problem Formulation}}
\toadd{Given a source C program $c$, an ideal translation is a \new{safe} Rust program $rs$ whose external behavior --- standard output, error messages, file modifications, and so on --- agrees with $c$ on every input in the safe universe $U$, defined as the set of inputs that do not trigger memory-unsafety in $c$. Formally, we want $rs$ to satisfy the Boolean requirement $\iseq{c}{rs}$, where $\iseqTxt$ is an oracle reporting equivalence over all of $U$. For inputs outside $U$, memory-safe handling is guaranteed by the \new{safe} Rust compiler itself.}


\toadd{However, existing LLM-based translators do not yet produce such an ideal translation. A standard setup provides translator $F$ with the C program $c$ and a small seed test set $T_0$, returning a Rust program $rs = F(c, T_0)$ that passes those tests. This ensures only the pointwise check ${\iseqTxt}^*(c, rs, t)$ on $t \in T_0$, where ${\iseqTxt}^*$ runs both programs on input $t$ and reports whether they match. On unseen $t \in U \setminus T_0$, ${\iseqTxt}^*(c, rs, t)$ may fail, so $rs$ does not satisfy $\iseq{c}{rs}$ over all of $U$ --- a generalization gap that we observe empirically (Sec.~\ref{sec:eval}).}
\toadd{Our aim in this paper is to develop a methodology $M$ that wraps the translator $F$ to narrow this gap. Given $rs = F(c, T_0)$, $M$ produces an improved translation $rs' = M(F, c, T_0)$ that agrees with $c$ on more unseen inputs than $rs$.}
%






\subsection{\fix{Adversarial Translation Loop}}
\label{sec:atl}

\toadd{We instantiate this methodology as an adversarial translation loop between two \fix{LLM} agents, inspired by GANs: a translator agent that produces a Rust translation, and a discriminator agent that searches for inputs on which the C and Rust programs disagree. The two agents work in turn, iteratively improving the translation, with the discriminator pushing the translator to fix newly exposed mismatches. The full pseudocode is given as Algorithm~\ref{alg:actor} in Appendix~\ref{sec:algodetail}.}

\toadd{Given the C program $c$, the seed test set $T_0$, a translator agent $F$, and a discriminator agent $D$, \our produces an improved Rust translation $rs'$:
\[
rs' = \text{ACToR}(F, D, c, T_0).
\]
$D$, the discriminator agent, surfaces mismatches via static inspection, manual testing, and a customizable differential fuzzer that the agent adjusts to fit each program's input space and efficiently explore inputs. Fuzzer construction is enabled by default; we ablate the no-fuzzer variant (\ournf) in Sec.~\ref{sec:ablation}. The translator $F$ is intentionally abstract: here we instantiate it as an LLM coding agent. \our can also post-process an existing Rust translation (e.g., output of C2SaferRust).}

\toadd{\our maintains an append-only test set $T$, initialized to the seed set $T_0$. \textbf{Initialization.} The translator produces an initial \new{safe} Rust translation $rs'$ that passes all of $T_0$ before the adversarial loop begins. \textbf{Discriminator step.} Reading $c$ and the current $rs'$, the discriminator emits a batch of tests on which their outputs disagree; each test is required to be well-formed and to execute successfully on $c$ before being appended to $T$. \textbf{Translator step.} The translator updates $rs'$ to pass the enlarged $T$ while \new{using only safe Rust}; if it cannot fully pass after a fixed number of retries, the loop continues with the latest $rs'$ retained, even when some tests in $T$ remain failing. After at most \textsc{MaxIter} iterations, \our returns the final $rs'$ and the accumulated test set $T$.}

\toadd{\our ensures the loop executes as expected through validation of $F$ and $D$, and file sandboxing. The translator's output $rs'$ is validated to contain no \texttt{unsafe} constructs and to pass all tests in the current $T$; on failure, the translator restarts within a retry limit. The discriminator's tests are validated to be well-formed, execute successfully on $c$, and disclose mismatches before being appended to $T$. Each agent additionally runs in a sandboxed workspace where its writes are effectively restricted to the files it controls (translator: Rust source; discriminator: tests), with $c$ and build scripts read-only to both. Implementation details of the validator and sandbox are deferred to Appendix~\ref{sec:impl}.}

\section{Evaluation}
\label{sec:eval}

\toadd{Ideally, a successful C-to-Rust translation preserves the C program's behavior on inputs unseen during translation, and produces fully safe Rust. We evaluate \our on the following criteria:}
\begin{enumerate}
    \item \toadd{\textbf{Correctness on unseen inputs.} Do \our's translations preserve the C program's behavior on inputs not seen during translation? Are the translations fully safe Rust?}
    \item \toadd{\textbf{Mechanism applicability.} Does the same adversarial loop transfer across different translator agents, LLMs, and program sizes?}
    \item \toadd{\textbf{Ablation.} How does the adversarial loop compare against alternative translators and alternative test-generation strategies? How sensitive are results to configuration choices?}
\end{enumerate}

\subsection{Experiment Settings}


\begin{table}[ht]
\centering
\caption{Benchmark Statistics}
\begin{subtable}{0.47\linewidth}
\centering
\caption{Micro Benchmark}
\begin{tabular}{|l|r|r|r|}
\hline
Program & LOC & Funcs & Cov. \\
\hline
printf & 371 & 11 & 95\% \\
expr & 452 & 17 & 92\%\\
fmt & 416 & 12 & 91\%\\
test & 528 & 17 & 78\%\\
join & 475 & 10 & 90\%\\
csplit & 303 & 7 & 88\%\\
\hline
\textbf{Total} & \textbf{2545} & \textbf{74} & \textbf{89\%} \\
\hline
\end{tabular}
\label{tab:micro_bms}
\end{subtable}
\hspace{0.01\linewidth}
\begin{subtable}{0.47\linewidth}
\centering
\caption{Macro Benchmark by Category}
\begin{tabular}{|l|r|r|r|}
\hline
Category & LOC & Funcs & Count \\
\hline
Text Processing & 13269 & 355 & 21 \\
File Dir.~Mgmt. & 6683 & 151 & 15 \\
Scripting Utils & 4204 & 118 & 8 \\
Sys.~Info Status & 2908 & 70 & 10 \\
Env.~Proc.~Ctrl. & 204 & 8 & 3 \\
\hline
\textbf{Total} & \textbf{27268} & \textbf{702} & \textbf{57} \\
\hline
\end{tabular}
\label{tab:macro_bms}
\end{subtable}
\label{tab:bms}
\end{table}



\textbf{Benchmarks.} 
Our evaluation uses two sets of benchmarks: a micro-benchmark set and a macro-benchmark set.
The micro-benchmark comes from a prior user study on C to Rust translation, which contains 6 standalone C programs, with an average of 424 lines of code each~\citep{ruishi2024ndss}.
We extend the validation suite to at least 60 test cases per program (89\% average line coverage), which are used to evaluate the correctness of translations.
We use this small-scale benchmark to demonstrate \our's ability to adapt to different LLMs and coding agent frameworks, and to perform an ablation study.

Our main macro-benchmark is a set of 57 real-world system utility programs (e.g., \texttt{ls}, \texttt{cp}, \texttt{xargs}) obtained from a software project named BSDCoreUtils~\citep{bsdcoreutil}. It has an average of \new{478} lines of code each.
\toadd{More details about how we filter the benchmark are reported in Appendix~\ref{filterbench}.}
\toadd{We use this benchmark to demonstrate \our's correctness improvement on real-world programs, and to compare its adversarial loop against alternative translators and the coverage-driven baseline.}
Given the large number of programs, manually crafting a comprehensive set of test cases for each macro-benchmark is infeasible,
and existing test suites for similarly-named GNU utilities are not directly applicable due to documented BSD/GNU behavioral differences.
Moreover, automated test generation techniques can produce test sets that unfairly advantage a specific translation approach. 
Instead, we perform a relative evaluation between \our and baselines, as described below in the Evaluation Metrics section.

To minimize measurement noise and make our experiments automatable, our evaluation focuses on single-threaded C programs that have deterministic behaviors across different executions, which is currently an assumption of our testing framework that both the translator and the discriminator depends on. 
Table~\ref{tab:bms} summarizes the statistics of two benchmarks, 
where we categorize the 57 programs into five categories: process execution control, file/directory management, shell scripting commands, system information access, and textual content processing.

\fix{
\textbf{Test Format and Testing Setup.}
As the adversarial procedure progresses, newly generated test inputs are continuously added to the test set.
To efficiently manage and execute these tests, we enforce a fixed test format and design a unified test runner that works for all benchmark programs.
Each test case is specified in a JSON format that defines the program input and the execution configurations. The configuration includes commands for test setup and cleanup, instructions for redirecting result files to standard output for comparison, and regular-expression rules to normalize outputs and filter out trivial differences (e.g., path names).
%
The test runner automatically executes the given set of tests for each program, runs both the C and Rust implementations with identical inputs, compares their outputs, and produces coverage reports, according to the provided command-line arguments.
}

\textbf{Agent Frameworks and Models.}
We use Claude Code~\citep{claudecode} with Claude Sonnet \new{4.5~\citep{sonnet45}} as the main agent framework and LLM for our experiments. 
To test if our approach generalizes to other agent frameworks and LLMs, we also performed experiments using Mini-SWE-Agent~\citep{miniswe}, a popular open-source agent framework, with \toadd{Claude 4.5 Sonnet, Claude 4 Sonnet, GPT 5 mini~\citep{gpt5mini}, Gemini-3.1-Pro~\citep{gemini3.1pro}, and one open source model, GLM-4.7~\citep{glm4.7}, as the LLMs.
We use `ClaudeCode+Sonnet4.5', `ClaudeCode+Sonnet4`, `SWE+Sonnet4.5', `SWE+Sonnet4', `SWE+GPT5mini', `SWE+GLM4.7`, `SWE+Gemini3.1-Pro` to represent them in later sections, respectively. The detailed version of the agent and model is reported in Appendix~\ref{modeldetail}.}

\textbf{Comparison Baselines.}
\toadd{We compare \our against four baselines:}
(1) \textit{Naive baseline:} a single translator agent identical to \our's, instructed to pass initially provided tests (seed tests in \our) without iterating to add tests or refine the code.
\toadd{(2) \textit{Strong baseline:} a single agent without iteration, but backed by a much stronger closed-source LLM (Claude Opus 4.6).}
\toadd{(3)} \textit{Coverage baseline:} a two-agent setup similar to \our, but the discriminator agent is only instructed to find additional tests that increase the line coverage of the C program, so the new tests do not specifically try to reveal inequivalent behavior.
\toadd{(4)} \textit{\ournf:} \our but without fuzzer construction.
\toadd{Comparison against (1), (3), and (4) isolates the gain from, respectively, \our's iterative loop, its adversarial test-generation strategy, and its agent-driven differential fuzzer; comparison against (2) calibrates these gains against a stronger underlying LLM.
We also check whether \our can improve the correctness of Rust translations produced by C2SaferRust, one of the recent works whose translations are public, compilable, and executable~\cite{nitin2025c2saferrust}.
}

\textbf{Evaluation Metrics.}
Our primary evaluation metric to evaluate the correctness of an individual translation is \textit{pass rate}, which is the percentage of tests that pass for a given set of tests.
%
%
\new{
On our micro-benchmark, we report the pass rate on manually written validation tests when we evaluate different configurations of \our only (e.g., different LLM–agent settings or experiment hyperparameters).
When we compare \our against other translation methods, we compute pass rates using a set of tests produced by a competing method, and refer to this metric as \textit{relative pass rate}.
This reduces bias from a method overfitting to its own tests (in practice, a method usually passes almost all of its own tests) and provides a stronger signal for cross-method comparison.
%
%
We use relative pass rate for (i) ablation studies that compare \our against alternative designs on the micro-benchmark, and (ii) the final experiment on the macro-benchmark, which lacks manually written tests and therefore relies solely on test suites generated by the methods.
%
The test sets used to compute relative pass rate achieve an average of \new{92.6}\% line coverage on the micro benchmark and \new{90.1\%} line coverage on the macro benchmark, by unioning test cases across translation methods.}
This indicates they are a useful signal for overall correctness as well.

\new{\textbf{Qualitative Requirements.}
%
\our aims to translate C code into fully safe Rust. Across both benchmarks, we enforced the absence of any \texttt{unsafe} blocks at every iteration.
After all the experiments finished, we manually reviewed the final translations output by \our, confirming they are indeed safe and contain no malicious constructs intended to bypass the checking.}
%

\textbf{Experimental Configuration.}
\new{We fixed a configuration to compare different agent settings and translation methods on the micro-benchmark.
By default,}
\our and all baselines start with 15 manually crafted, diverse seed tests based on the C code for each program\footnote{The seed tests carry intended test format that both translator and discriminator of \our use. \new{It also avoids a cold-start that produces translations in poor quality at the first few iterations} for \our as well as all other systems we compare to.}.
For the coverage baseline, \ournf, and \our, we run 10 iterations of the adversarial loop. On each iteration, we generate 3 new tests, resulting in a total of 45 tests per program in the end.
%
\new{We fix these parameters because exploring different configurations for all agent–model settings and translation methods would be prohibitively costly.}
\new{We conduct an ablation experiment to study the influence of different experiment configurations on the final correctness and cost of \our,
and apply the default configuration to translate the macro-benchmark.}
See the Appendix~\ref{sec:prompt} for the prompt templates.

\subsection{Evaluation Result on Micro Benchmark}
\label{sec:ablation}
\begin{figure*}[th!]
\centering
\includegraphics[width=0.8\textwidth]{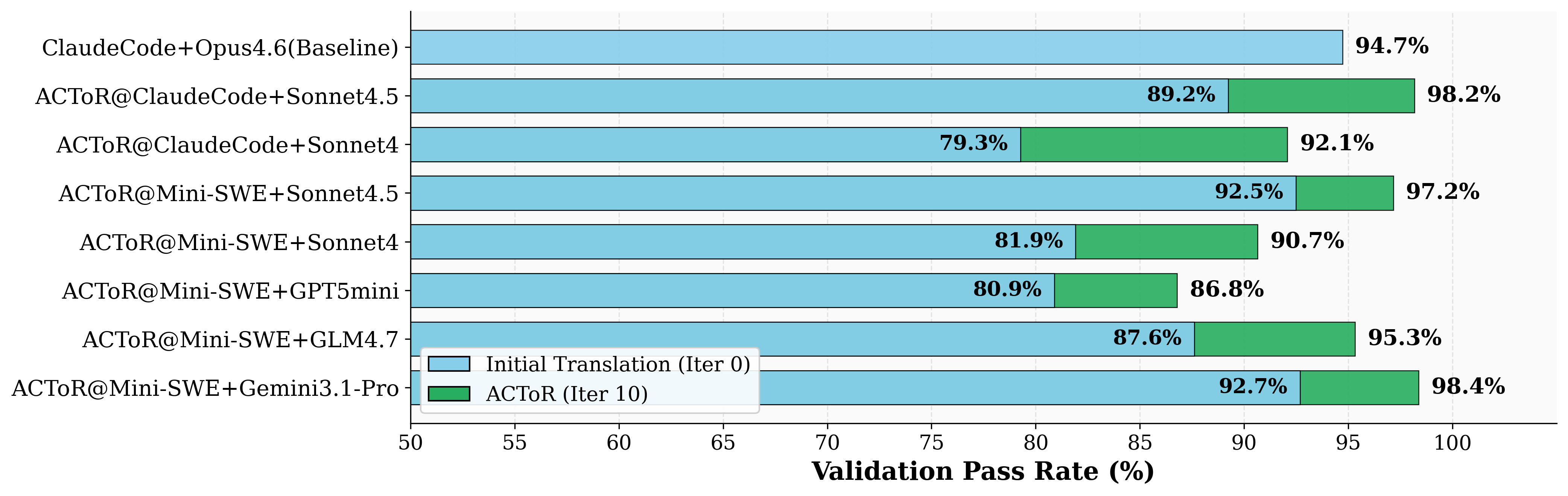}   
\caption{
\toadd{Overall pass rate (on hidden tests) on the micro benchmark across 7 agent--LLM configurations, comparing \our (green, iter 10) with the naive baseline (cyan, initial translation) and the strong baseline (Claude Opus 4.6, no iteration).}}
\label{fig:res1}
\end{figure*}

\textbf{Correctness \& Adaptability.}
\toadd{We compare \our against the \textit{naive baseline} (per agent--LLM configuration) and the \textit{strong baseline} (Claude Opus 4.6).}
%
%
The aggregate results are shown in Figure~\ref{fig:res1}; \toadd{per-program results are provided in Appendix~\ref{perprogram}.}
All translations from \our and the baselines are free of \texttt{unsafe} blocks.
\our consistently 
improves the translation correctness on all \new{7} agent-model settings.
With Claude Code \new{and Claude-Sonnet-4.5}, the validation 
pass rate rises from an average of \new{89.2}\% (naive baseline) to \new{98.2}\% after 10 iterations.
With Mini-SWE-Agent, completing 10 iterations improves correctness by \toadd{4.7\% to 8.8\%} across the five LLMs.
\toadd{Against the strong baseline (94.7\%), \our with Claude-Sonnet-4.5 (98.2\%) is higher despite a less capable LLM. This suggests that the iterative refinement provides value beyond what is gained from using a stronger underlying model alone. The consistent gains across all seven configurations further indicate the loop transfers across underlying translators.}
\new{
We use Claude Code + Sonnet-4.5 as the default configuration for the following experiments.
With this configuration, it takes 411 million tokens to finish 6 programs, which costs \$201 USD in total. The average time cost for \our to finish one program is approximately 2.45 hours\footnote{The token consumption is computed by summing up the input, output, and cached tokens reported in the API response messages. When computing the time cost, we first compute the average time cost per iteration, then times it by 10 as an approximation of the total runtime for translating one program.}.}
%


%

\begin{figure}[t]
    \centering

    \begin{subfigure}[t]{0.495\textwidth}
        \centering
        \includegraphics[width=\linewidth]{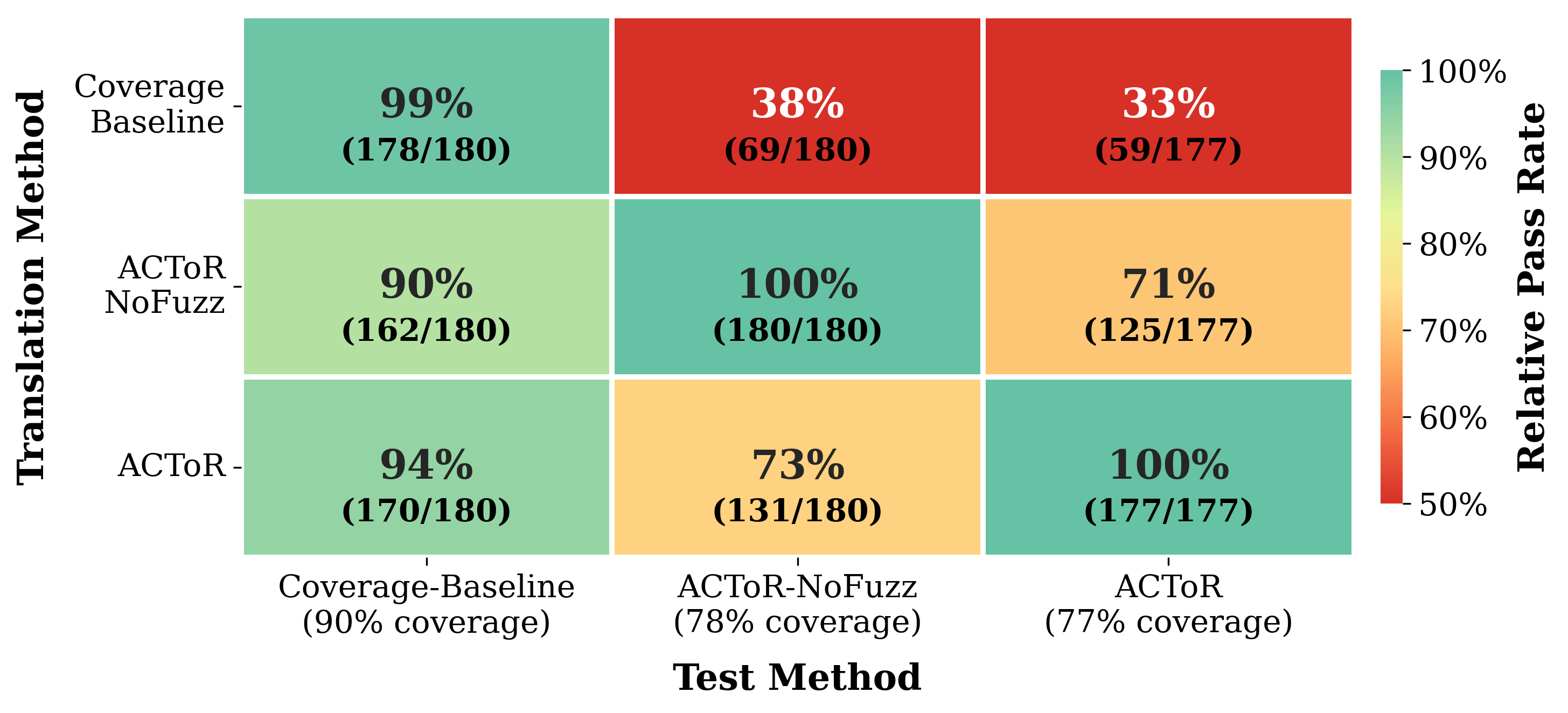}
    \end{subfigure}
    \hfill
    \begin{subfigure}[t]{0.495\textwidth}
        \centering
        \includegraphics[width=\linewidth]{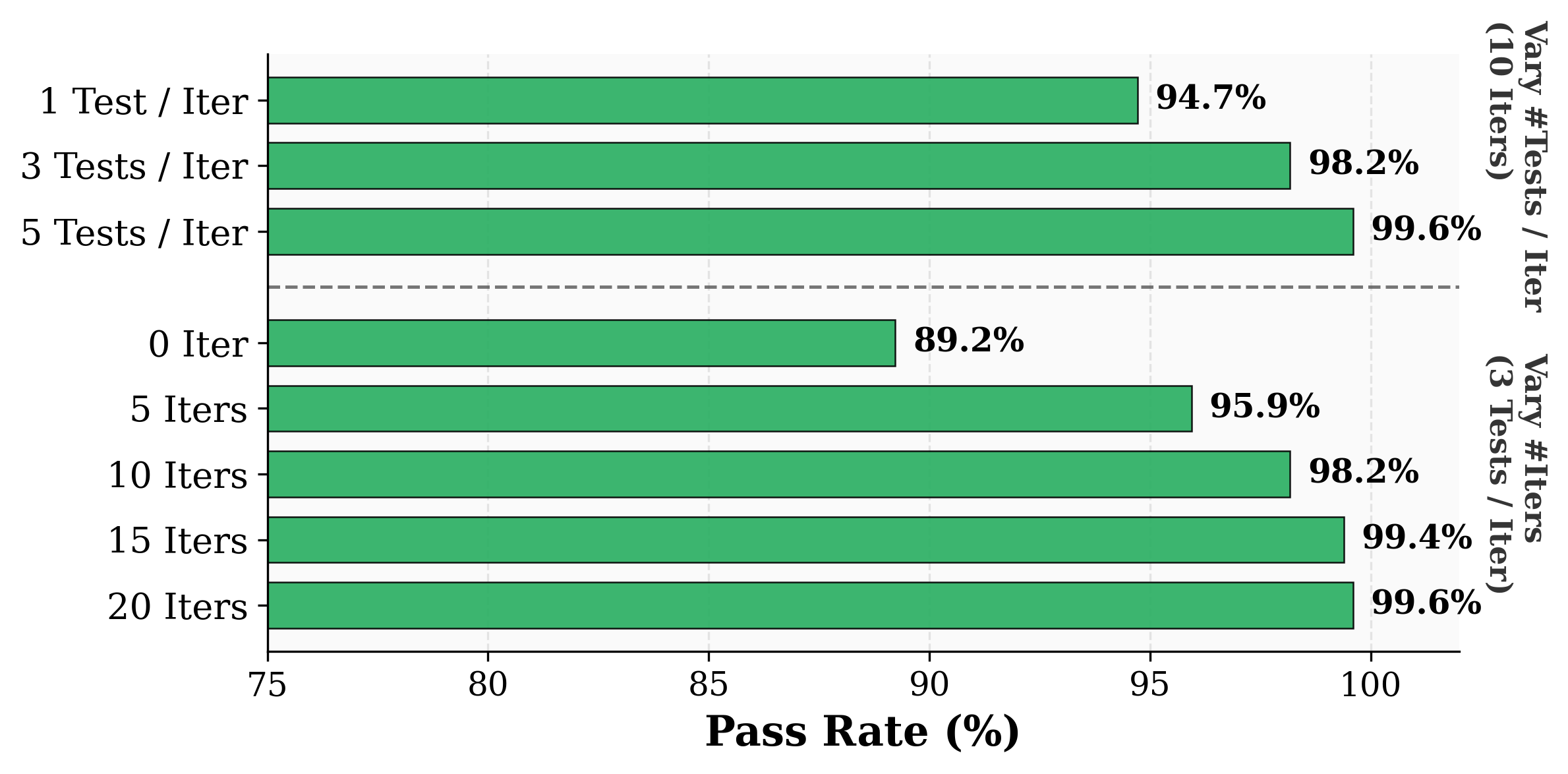}
    \end{subfigure}

    \caption{(Left) The relative comparison among 3 translation methods on Claude Code with Sonnet-4.5. Entry $(\text{row}, \text{column})$ is the relative pass rate of the row's translation on the column method’s tests. (Right) The validation pass rate of \our on different configurations. Pass rate versus the number of new test cases added per iteration (Top). Pass rate versus the number of iterations (Bottom).}
    \label{fig:ablation2}
\end{figure}

\textbf{Ablation Study on Design.}
We conduct an ablation to know how the adversarial design in \our drives correctness gains on the micro-benchmark.
%
\toadd{We compare three variants of \our---coverage-baseline (replacing the adversarial discriminator with non-adversarial test generation), \ournf (without the fuzzing script), and the full \our---to isolate the contributions of the \textit{adversarial discriminator} and the \textit{use of the fuzzing script}.}
\toadd{
The cross-testing results (shown on the left of Figure~\ref{fig:ablation2}) report the relative pass rate of each method (row) on the tests generated by the other methods (column).
While all methods pass most of their own tests, the adversarial variants generalize much better across tests.
On coverage-guided tests, \ournf and the full \our achieve 90\% and 94\% pass rates, respectively, whereas the coverage-guided baseline achieves less than 40\% on tests from either adversarial variant.
This indicates that adversarially generated tests are more effective at exposing semantic mismatches than coverage-guided sampling.
%
The full \our further improves over \ournf in cross-testing \toadd{(71\%$\rightarrow$73\%)}, suggesting that the customized fuzzer helps the discriminator find more challenging bugs.
To better understand cost-effectiveness, we also run an \textit{equal-cost} experiment in which we extend the coverage baseline to 25 iterations, roughly matching the cost of running \our for 10 iterations.
However, even under equal budget, the coverage baseline's pass rate \toadd{(75/177 = 42.4\%)} remains far below that of \our \toadd{(425/447 = 95.1\%)}.
%
%
More details about this ablation study and equal-cost experiment are reported in Appendix~\ref{app:ablation-design}.}

\textbf{Ablation Study on Configurations.}
\toadd{We also study how different configurations affect \our on the micro-benchmark. We examine three factors: \textit{seed tests}, \textit{number of iterations}, and \textit{number of tests} generated by the discriminator agent per iteration.}
\toadd{
Details are in Appendix~\ref{app:ablation-config}.
%
Figure~\ref{fig:ablation2} (right) shows the results.
The results show that the default configuration, which uses 15 initial seed tests, 10 iterations, and 3 new tests per iteration, achieves a favorable balance between performance and cost.
%
Increasing tests per iteration yields diminishing returns: 3 tests instead of 1 improves pass rate from 94.7\% to 98.2\% with a 16\% cost increase, while 5 tests adds only 1.4\% at an additional 29\% cost.
Most gains occur within the first 10 iterations; extending to 20 iterations adds only 1.4\% at 2.1$\times$ cost.
Finally, using 15 seed tests instead of a single seed test improves pass rate from 94.9\% to 98.2\% with only a 6\% cost increase, suggesting that richer initial coverage is beneficial.
}

\new{\textbf{Stability of Results.}
To assess the stability of \our’s translation results, we run \our three times using the default settings (i.e., 15 initial tests, 10 iterations, and 3 tests per iteration). We record the pass rate of each trial on the validation set. For each program, we compute the \textit{sample standard deviation} of the three pass rates and then average these standard deviations across all programs. On average, the three runs achieve a pass rate of 97.0\%, with a deviation of 1.9\%.
}

\graybox{
\toadd{On the micro-benchmark, \our substantially and stably improves translation correctness over the non-adversarial baselines.} The adversarial discriminator is crucial to improving correctness, and the fuzzer construction further strengthens the discovery of mismatches.
}


\begin{figure*}
\centering
\includegraphics[width=0.99\textwidth]{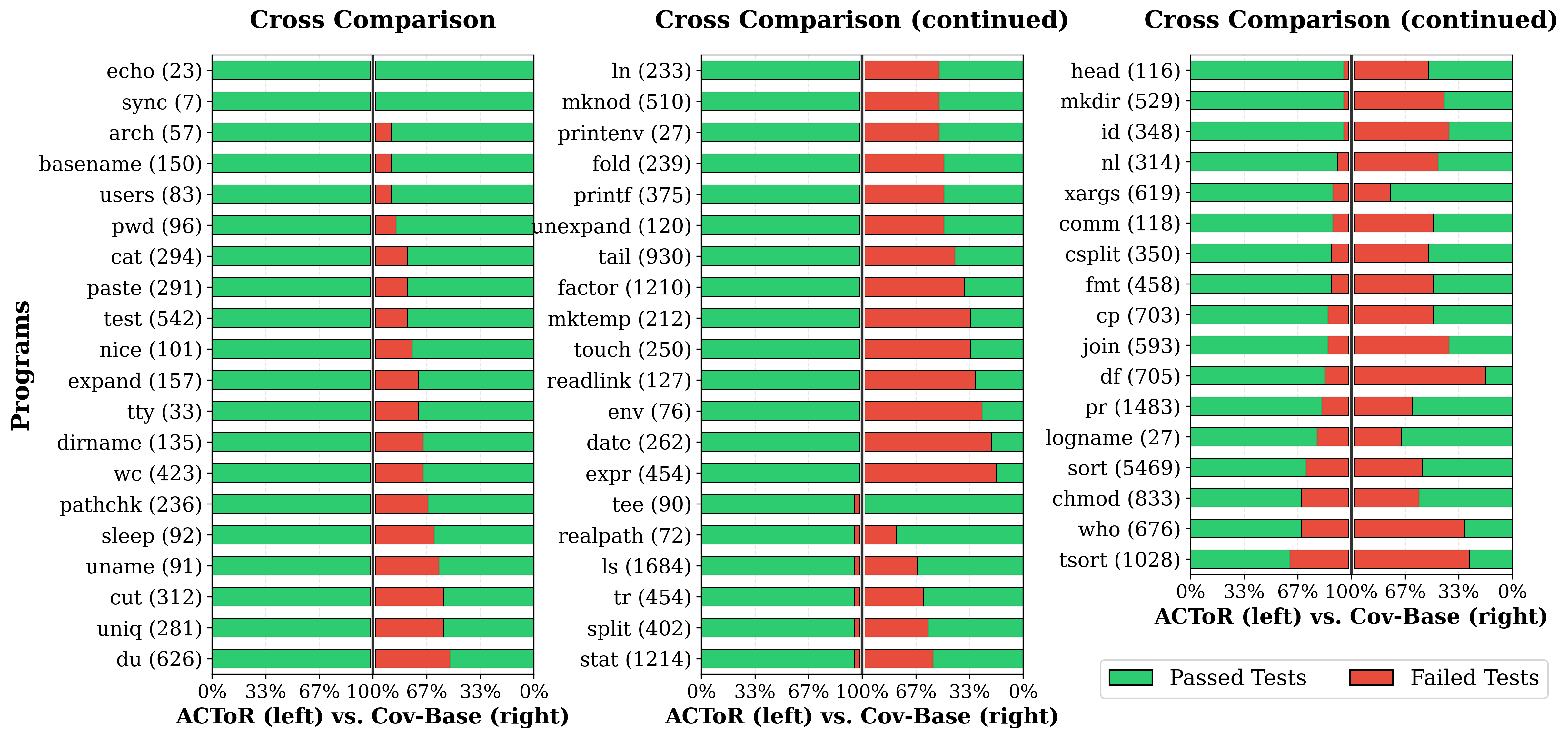}   
\caption{
On the macro benchmarks, the relative pass rate when cross-comparing \our and coverage baseline (Cov-Base) on Claude Code with Claude-Sonnet-4.5 at iteration 10. For each program, the left bar shows evaluating the translation from \our on tests generated in coverage-baseline; the right bar is evaluating the translation from coverage-baseline on \our's tests. The length of each program in LoC is presented next to the program name.
}
\label{fig:macro-res-new}
\end{figure*}
\subsection{Evaluation on Macro Benchmark}\label{sec:macrores}

\toadd{We compare \our against two baselines on the macro benchmark: the coverage baseline (using \textit{relative pass rate}) and the strong baseline (Claude Opus 4.6, evaluated on the coverage baseline's tests for a fair comparison with \our).}
We use the Claude Code agent and \new{Claude-Sonnet-4.5} for this experiment\footnote{We run macro evaluation on both Claude-Sonnet-4 and Claude-Sonnet-4.5. Due to space limitations, we put the result of Claude-Sonnet-4 in Appendix~\ref{sec:oldmacro} for reference. The results are consistent between two trials.}.
%
\new{All \our translations are free of \texttt{unsafe} blocks.}

\toadd{Figure~\ref{fig:macro-res-new} shows the cross-comparison at iteration 10.}
The left bar for each program is the relative pass rate of testing \our's translation on the test cases from the coverage baseline. The right bar shows the reverse: relative pass rate of the coverage baseline against \our's test cases.
%
\toadd{\our outperforms the coverage baseline on \new{54/57} programs.}
\our gets 100\% relative pass rate on \new{34} programs when running on the tests from the coverage baseline, while the coverage baseline can only get full pass on \new{3 programs}.
%
\toadd{On average, \our achieves 95.1\% relative pass rate, 36.7\% higher than the coverage baseline's 58.4\%.}
The cross-comparison confirms that adversarial agentic design in \our plays a significant role in improving the translation correctness.
%
\toadd{As an absolute measure}, by taking the union of the test cases from both \our and the coverage baseline, we get a test set of an average \new{90.1}\% coverage.
\our achieves a \new{95.9}\% pass rate on it.
%
%
\toadd{This shows \our's utility on real-world projects.}
\new{The total financial costs to translate 57 programs using the coverage baseline and \our are \$808 USD and \$1634 USD, respectively. On average, \our consumes 57 million tokens per program, with more than 97\% of them being cache-read tokens.
The average time cost for \our to finish one program is about 2.5 hours. 
}
%


\toadd{Against the strong baseline (Claude Opus 4.6), we compare \our and the strong baseline on the coverage-baseline tests (neither method sees them during translation). Across the 57 programs, \our wins 32, the strong baseline wins 3, and the two tie on 22 programs where the tests cannot distinguish them. On the 35 discriminating programs, \our wins 32 (91\%), confirming the micro-benchmark pattern at scale: the gain from \our's adversarial loop is not closed by simply switching to a stronger underlying LLM.}

\toadd{\textbf{Augmenting C2SaferRust.} We also test \our as a correctness-improvement layer on top of C2SaferRust~\citep{nitin2025c2saferrust}, one recent C-to-Rust translator. From C2SaferRust's benchmark, we filter out libraries that are not independently testable, leaving \todo{7} standalone executable programs within \our's scope, and seed \our with C2SaferRust's translation as the initial Rust input. Because C2SaferRust's output contains \texttt{unsafe} blocks, we relax \our's strict safe-Rust requirement for this experiment: the translator agent may keep existing \texttt{unsafe} code but is instructed not to introduce additional \texttt{unsafe} lines. On a held-out validation test suite, applying \our raises pass rate by approximately \todo{16.6\% (from 76.3\% to 92.9\%)} over C2SaferRust's translations alone, showing that \our can complement existing translation techniques to further improve correctness (Sec.~\ref{sec:atl}).}

\graybox{
\toadd{\our improves correctness on the BSDCoreUtil benchmark: it outperforms the coverage-driven baseline on 54/57 programs (95.1\% vs.\ 58.4\% relative pass rate) and the strong baseline (Claude Opus 4.6) on 32/57 programs with 22 non-discriminating ties. As a layer on top of C2SaferRust, \our helped raise the pass rate on hidden tests from \todo{76.3\% to 92.9\%}.}
}

\subsection{\toadd{Limitations}}
\label{sec:limitations}
\toadd{\our currently targets standalone command-line programs and does not directly support library translation, due to the lack of a precise equivalence notion when program behaviors are language-specific---known to be hard to automatically determine~\citep{matchfixagent}, especially when library interfaces and types differ across languages. Existing work targeting C-to-Rust library translation often addresses this with manually vetted test drivers, one written for each language to drive the library's interface (e.g., CRUST-Bench~\citep{anirudh2025crust}); integrating such test-driver synthesis with our discriminator's adversarial fuzzing is left to future work. Our macro evaluation uses closed-source models. However, the micro-evaluation has shown that the adversarial loop generalizes across multiple agent--LLM configurations with consistent gains, suggesting the same pattern would hold across other model families.}

\section{Conclusion}
\label{sec:conclusion}

We present \our, an agent-based automatic translation approach that aims for functionally correct C to Rust translations.
Our method introduces a collaborative pipeline in which two agents iteratively refine the translations.
At each iteration, the discriminator agents detect the semantic mismatches between the source code and the translated program, and report them as failing test cases.
The translator agents then improve the translation to pass the current test set.
Our experiments demonstrate the power of \our's adversarial agentic setup to produce translations with high correctness, surpassing the quality of translation as compared to non-adversarial setups.
\toadd{A noteworthy advantage of \our is that its adversarial loop is largely independent of the underlying translator and LLM.
ACToR also composes with existing translators: applied on top of C2SaferRust's outputs, it further improves the pass rate on hidden tests.
}

\bibliographystyle{plainnat}
\bibliography{neurips}
\appendix







\section{\toadd{Adversarial Translation Loop: Detailed Algorithm}}
\label{sec:algodetail}

\toadd{\our takes as input the C code $c$, the universe $U$, the seed test set $T_0$, together with hyperparameters such as the number of iterations the translator and discriminator interact (MaxIter) before \our outputs the final translated program. \our then conducts iterative translation on the source C code and outputs both the Rust translation $rs^{*}$ and the final test set $T^{*}$. The \textsc{Translator} subroutine in Algorithm~\ref{alg:actor} is intentionally abstract: in our default instantiation it is an LLM coding agent. To complement an existing C-to-Rust translator, the loop can also be seeded with that translator's output (e.g., C2SaferRust), with \our then post-processing the output across iterations.}

\toadd{During iterative improvement, \our maintains an append-only test set $T$. In the beginning, the translator agent generates the initial \new{safe} translation and ensures it passes {\em all the seed tests} (lines 1-3) before the adversarial iterative loop (involving the discriminator) begins. Next, in each iteration that follows, the discriminator agent reads the C code $c$ and the current Rust code $rs$, and generates test cases on which C and Rust outputs disagree (lines 7-11). To efficiently surface such inputs, the discriminator constructs and refines a differential fuzzer besides static inspection and manual testing.}

\toadd{The discriminator is structured so that (1) all the test cases abide by the correct format and execute successfully on the source C program (SANITYOK at line 9); and that (2) some test cases witness a mismatch between outputs of the C and the Rust code that the translator has created so far (line 10). We ensure this by providing a prompt to the discriminator to continue retrying until the generated test cases meet the above two requirements. Once successful, the newly generated test cases ($Batch$) are added to the full test set. The translator agent then takes the updated test cases as guidance to improve the translation. It is set up with prompts that repeatedly ask it to pass all the test cases in the current test set and \new{only use fully safe Rust} (lines 13-16). If the translator runs out of retries but still cannot fully pass the test set $T$, the loop continues with the latest $rs$ retained, even when some tests in $T$ remain failing. After a fixed maximum number of iterations, \our stops the loop and outputs the final translation and the test set generated.}

\begin{algorithm}[t]
\caption{ACToR: Adversarial C to Rust Translation}
\label{alg:actor}
\begin{algorithmic}[1]
\Require C program $c$, seed test set $T_0$, test input universe $U$ (implicit)
\Statex\hspace*{-\algorithmicindent}\textbf{Params:} MaxIter (of outer loop), TestBatchSize, RetryLimit (of inner loops)
\Ensure Final translation $rs^*$ and final test set $T^*$

\While{\textsc{Not}\Call{PassesSuite}{$c, rs, T_0$}} \Comment{initial translation on seed tests}
    \State $rs \gets$ \Call{Translator}{$c, \varnothing, T_0$}
\EndWhile
\State \textbf{assert} \Call{IsSafe}{$rs$};\quad $T \gets T_0$

\For{$k=1$ \textbf{to} MaxIter}
  \State $retry \gets 0$
  \Repeat
    \State $Batch \gets$ \Call{Discriminator}{$c, rs, n{=}\text{TestBatchSize}$};
    \State $IsValid \gets \forall t \in Batch:\; \Call{ValidTestFormat}{t} \land \Call{SanityOk}{t, c}$
    \State $IsDisc \gets \exists t \in Batch:\; \neg \iseqTxt^*(c,rs,t)$;\quad $retry \gets retry + 1$
  \Until{$(|Batch|=\text{TestBatchSize} \land IsValid \land IsDisc) \lor retry \geq \text{RetryLimit}$}
  \State $T \gets T \cup Batch$;\quad $retry \gets 0$
  \While{\textsc{Not}~\Call{PassesSuite}{$c, rs, T$} $\land\; retry \leq \text{RetryLimit}$}
    \State $rs \gets$ \Call{Translator}{$c, rs, T$}; \quad $retry \gets retry + 1$
  \EndWhile
  \State \textbf{assert} \Call{IsSafe}{$rs$}
\EndFor
\State \Return $rs$ as $rs^*$, $T$ as $T^*$
\end{algorithmic}
\end{algorithm}

\section{ACToR Implementation}
\label{sec:impl}

\fix{
The ACToR translation pipeline consists of three components: a task manager, a translator agent, and a discriminator agent.
The task manager prepares a sandboxed working directory containing all necessary input files for the target project, coordinates agent execution, and snapshots the workspace after each iteration.
Both agents operate exclusively within this sandbox.
To enforce isolation, each agent is restricted to modifying only the files under its control: the translator agent may modify only the translated Rust source file, while the discriminator agent may modify only the test case files.
All other files, including the original C source code, documentation, and build scripts, are read-only for both agents.
This constraint is enforced indirectly: after each iteration, all files outside the allowed edit list are reverted to their state from the previous iteration.
The agents do not communicate via explicit message passing; instead, they collaborate implicitly through the shared filesystem state in the sandbox.
This design simplifies coordination and allows agents to be easily replaced without affecting other components.
The task manager alternates execution of the two agents in an adversarial loop.
At iteration 0, the translator agent produces an initial Rust translation.
Subsequently, the discriminator and translator alternate as
$[\text{discrimination} \rightarrow \text{translation}]_{\text{iter}_1} \rightarrow [\text{discrimination} \rightarrow \text{translation}]_{\text{iter}_2} \rightarrow \ldots$
This process continues for a fixed, configurable number of iterations.}

\fix{During the \emph{translation phase}, the translator agent may execute all existing test cases, analyze any observed mismatches by inspecting both the C source and the Rust translation, then repair the translated code.
The agent must ensure that the updated Rust program passes all tests before completing the task.
Upon submission, an independent validator function will be invoked to verify that (i) the Rust code compiles successfully, (ii) the translation contains no \texttt{unsafe} constructs, and (iii) all tests pass.
If any of these conditions are violated, the translator agent is restarted to repeat the task.
The prompts used for the translator agent are provided in Appendix~\ref{sec:prompt}.}

\fix{During the \emph{discrimination phase}, the discriminator agent analyzes the C program and its Rust translation to identify potential behavioral discrepancies.
It then refines the fuzzer to explore a large input space.
After confirming mismatches, the agent selects the top-$k$ distinguishing inputs and formats them as new test cases.
A validator checks that exactly $k$ new tests are added and that the Rust translation exhibits mismatches with the C program on these inputs.
To filter out discrepancies caused by non-determinism unrelated to translation correctness, the checker also compiles the C source under two different compilation configurations (e.g., distinct optimization levels) and verifies that both binaries produce identical outputs on each test input.
The discriminator agent retries until all checks pass or a retry limit is reached.
The prompts used for the discriminator agent are also provided in Appendix~\ref{sec:prompt}.}

\toadd{Each agent runs in a sandboxed workspace where the modification is restricted to the files it controls (translator: Rust source; discriminator: tests), with C code and the build scripts read-only to both. This is enforced indirectly: after each iteration, any file outside the agent's allowed edit set is reverted to its prior content.}

\fix{
ACToR system is designed to be lightweight: the worker manager is implemented in around 1.8k lines of Python code, while each agent is implemented with less than 500 lines including the prompts.
}

\section{Agent and Model Details}
\label{modeldetail}
We use Claude Code version 2.0.42 in all experiments. The model versions are claude-sonnet-4-5-20250929, claude-sonnet-4-20250514, gpt-5-mini-2025-08-07, gemini-3.1-pro-preview, and glm-4.7.

\section{Benchmark Filtering}
\label{filterbench}
The full set of BSDCoreUtils contains 69 programs, but we filter out 12 of them from our main results due to three reasons: (1) some utilities are trivial (e.g., \code{true}, \code{yes}) (2) certain utilities require special environments (e.g., \code{stty}) or privileges (e.g., \code{sudo}) not available in our experimental sandbox; and (3) certain utilities perform destructive operations (e.g., \code{rmdir}) that may break the testing harness.

\begin{figure}[t]
\centering
\begin{tcolorbox}[promptbox, title= (\our / Coverage Baseline) Translator Agent Prompt]
\begin{lstlisting}[language=Markdown, basicstyle=\ttfamily\tiny]
Task Description:
```
You are an expert in C and Rust.
Your task is to translate the C `<project_name>` project to safe Rust implementations.

<...Project Setup>

## Workflow
1. Read the C code and the test script to understand the functionalities.
2. Initialize a new Cargo project in the `ts/` folder.
3. Translate the C code to Rust code and compile it into binary.
4. Run `./testcmp.sh` to compare the output of the translated Rust code with the original C program. You should run `./testcmp.sh --help` to understand how to use the test script.
5. Clean the working directory by removing temporary files and backup files.

## constraints
- The translated Rust code MUST compile and MUST be 100% safe.
- The translated Rust code MUST pass all the unit tests.
```
\end{lstlisting}
\end{tcolorbox}

\caption{The task prompt for the translator agent.}
\label{fig:promptmacro-1}
\end{figure}

\begin{figure}[t]
\begin{tcolorbox}[promptbox, title= (Coverage Baseline) Test Generator Agent Prompt --- Adding Tests]
\begin{lstlisting}[language=Markdown, basicstyle=\ttfamily\tiny]
Task Description:
```
You are an expert in C and Rust.  
Your task is to add additional test cases to for the C program to improve the coverage.

<...Project Setup>

## Workflow
1. Read the C code to understand the functionalities.  
2. Focus first on **core functionalities**, then explore **edge cases**.  
3. Run `make clean && make all && ./testcmp.sh coverage` to compile the C code and get the current coverage.
4. Read the coverage report and the record of added test cases in `test_cases_record.md` to find potential missed cases.
5. Design **3** new test cases that are different from existing test cases.
6. Run `./testcmp.sh coverage` to get the new coverage. Ensure that the new coverage is higher than the previous one.
7. Clean the working directory by removing temporary files and scripts, temporary test cases, and backup files.

<...Test formatting requirements>

## Constraints
1. There should be exactly 3 new test cases added to the JSONL file. You should run `./testcmp.sh` and the number of test cases will be shown. There should be `<current_test_cases_number> + 3` test cases in total.
2. The 3 test cases should be different from each other. You should check this by reading the content of the test cases.
3. The added tests must be valid for the C code. You should run `make clean && make all` and then run `./testcmp.sh compare ./xxx.out(compiled from C code)`. It must show `All tests passed!`.
```
\end{lstlisting}
\end{tcolorbox}

\caption{The task prompt for the test generator agent of the coversge baseline.}
\label{fig:promptmacro-2}
\end{figure}

\begin{figure}[t]
\begin{tcolorbox}[promptbox, title= (\our) Discriminator Agent Prompt --- Adding Tests]
\begin{lstlisting}[language=Markdown, basicstyle=\ttfamily\tiny]
Task Description:
```
You are an expert in C and Rust.  
Your task is to add additional test cases to discover semantic mismatches between the C code and the translated Rust code.

<...Project Setup>

## Workflow
1. Analyze the C code and the translated Rust code to detect **semantic mismatches**.  
2. Focus first on **core functionalities**, then explore **edge cases**.  
3. Read the current test script and the record of added test cases in `test_cases_record.md` to find potential missed cases.
4. Collect the best **3** new input cases that expose mismatches between C code and Rust translation. Add the 3 new test cases to the test cases file.
5. Run the new tests to compare the output of the translated Rust code with the original C program to confirm the mismatches.
6. Clean the working directory by removing temporary files and scripts, temporary test cases, and backup files.

<...Test formatting requirements>
IF {$allow_fuzz}
<...Details on the usage of the fuzzer script>
ENDIF

## Constraints
1. There should be exactly 3 new test cases added to the JSONL file. You should run `./testcmp.sh` and the number of test cases will be shown. There should be `<current_test_cases_number> + 3` test cases in total.
2. The 3 test cases should be different from each other. You should check this by reading the content of the test cases.
3. The added tests must be valid for the C code. You should run `make clean && make all` and then run `./testcmp.sh compare ./xxx.out(compiled from C code)`. It must show `All tests passed!`.
4. The added tests should reflect the differences between the C code and the Rust code. You should run `./testcmp.sh compare ./ts/target/release/xxx(compiled from Rust code)`. The Rust code should fail on all 3 new test cases.
```
\end{lstlisting}
\end{tcolorbox}

\caption{The task prompt for the discriminator agent of \our.}
\label{fig:promptmacro-3}
\end{figure}



\begin{figure}[t]
\begin{tcolorbox}[promptbox, title= Post-processing Prompt --- Eliminating Unsafe Blocks]
\begin{lstlisting}[language=Markdown, basicstyle=\ttfamily\tiny]
Task Description:
```
You are an expert in C and Rust.
Your task is to translate the C `<project_name>` project to safe Rust implementations.

<...Project Setup>

## Workflow
1. Read the C code and the test script to understand the functionalities.
2. Initialize a new Cargo project in the `ts/` folder.
3. Translate the C code to Rust code and compile it into binary.
4. Run `./testcmp.sh` to compare the output of the translated Rust code with the original C program. You should run `./testcmp.sh --help` to understand how to use the test script.
5. Clean the working directory by removing temporary files and backup files.
```

The translation contains unsafe keywords.
Please fix it to ensure no unsafe code is used.
Also, keep it passing all the tests.
\end{lstlisting}
\end{tcolorbox}

\caption{The task prompt for eliminating unsafe blocks.}
\label{fig:promptpost}
\end{figure}

\section{Prompt Template}
\label{sec:prompt}
For Mini-SWE-Agent, we keep the system prompt that describes the agent workflow unchanged as the original one during all the experiments.
For Claude Code, we begin directly with the task described in the user message, without any additional text preceding it.
The task prompt structure for the translator agent is shown in Figure~\ref{fig:promptmacro-1}. This prompt is shared among all translation methods during evaluation.
The prompt structure for the test-generation agent used in the coverage baseline is shown in Figure~\ref{fig:promptmacro-2}.
The prompt structure for the discriminator agent for \our is shown in Figure~\ref{fig:promptmacro-3}. The fuzzing mode is controlled by the `allow\_fuzz` section in the prompt.


\begin{figure*}[th!]
\centering
\includegraphics[width=0.97\textwidth]{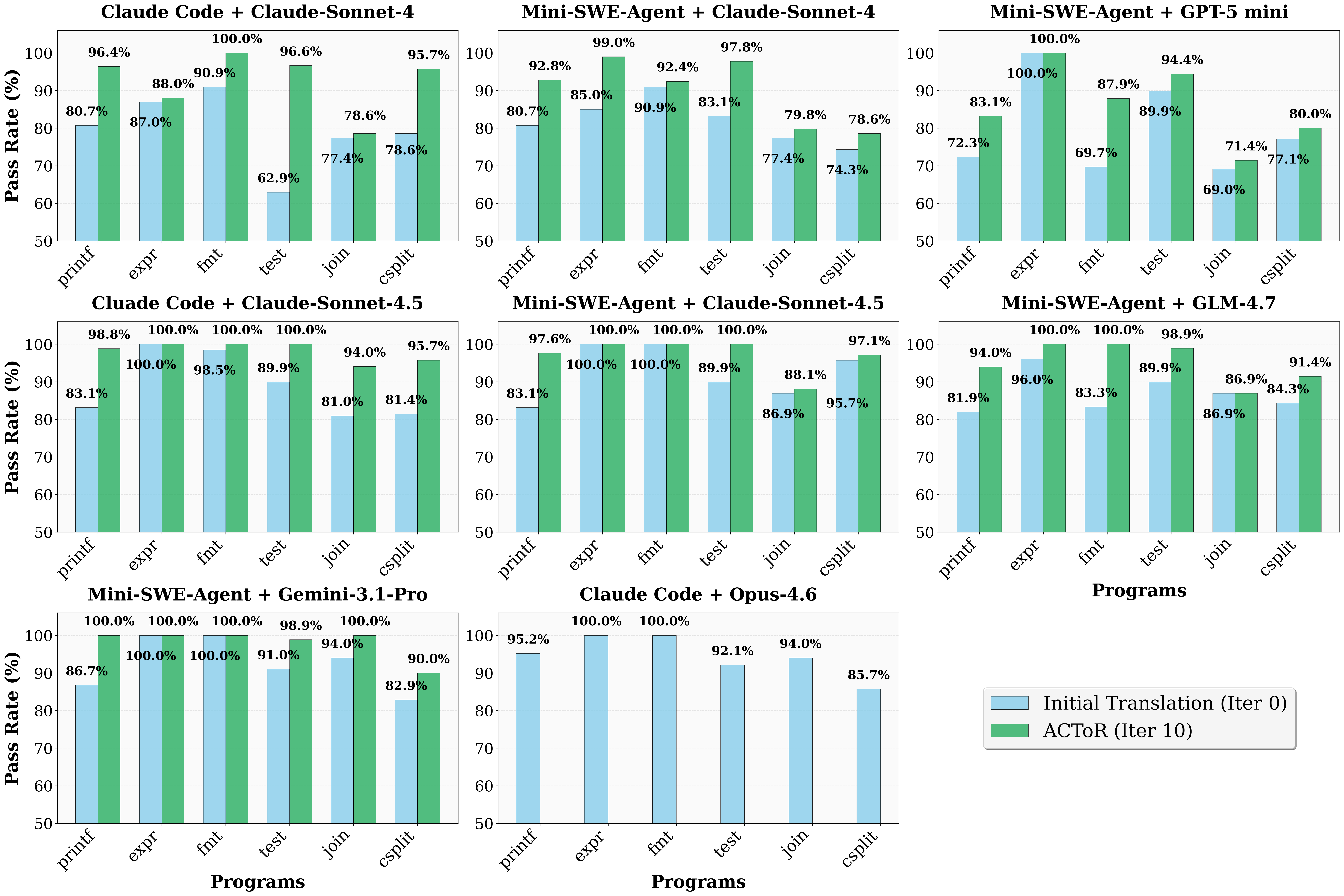}   
\caption{
Validation pass rate achieved by \our on micro benchmark across different settings compared with naive baseline and strong baseline, on \new{7} different agent-model settings.}
\label{fig:res1-perprogram}
\end{figure*}

\section{Per-program Results for Correctness \& Adaptability.}
\label{perprogram}
The per-program results of the Correctness \& Adaptability experiments are shown in Figure~\ref{fig:res1-perprogram}.
When applying SWE + GLM-4.7 to the translation loop, we were unable to complete 10 iterations for the program `join` within 48 hours due to limited model-serving capacity. We therefore report results from the maximum completed iteration on this program. Completing the full 10 iterations may yield larger correctness improvements.

\section{Ablation Study on Design}
\label{app:ablation-design}
We conduct an ablation to know how the adversarial design in \our drives correctness gains on the micro-benchmark. 
We analyze the improvement offered by two design choices:
(a) having an \textit{adversarial discriminator} in the workflow to guide the translator, and (b) forcing the discriminator agent to use the \textit{fuzzing script} we provided.
We compare three variants of \our: coverage-baseline, \ournf, and the full \our (with fuzzing).
%
%
%
The results summed across programs at the final iteration are reported on the left of Figure~\ref{fig:ablation2}. Entry $(\text{row}, \text{column})$ shows the relative pass rate of the row method on the column method’s tests.
All three methods largely pass their own tests as expected.
On coverage-baseline's tests, \ournf and the full \our reach \toadd{90\% and 94\%} passing rates, respectively; in the reverse direction, coverage-baseline achieves less than \toadd{40}\% passing rate on tests from either \our variant. 
This indicates that adversarial discriminators produce tests that more effectively expose semantic mismatches and result in more correct translations.
Between \our variants, \toadd{the full \our attains 73\% vs.\ 71\%} when cross-tested against \ournf. This suggests that the discriminating agent uses the \fix{customized fuzzer} to surface more challenging bugs in the translated code.
%
We also measure the final test coverage of each method on the C code. The result is shown at the bottom left of Figure~\ref{fig:ablation2}.
%
The coverage baseline works as expected and achieves the highest line coverage on the C code than two other \our variants. However, it is worse in correctness. This further supports the conclusion that adversarial translation is more efficient in improving correctness compared with purely sampling new test cases based on coverage.
\new{Finally, we compare the financial costs of completing all six programs: the coverage baseline costs \$95 USD, while \ournf and the full \our cost \$240 USD and \$211 USD, respectively.}
To better understand cost-effectiveness, we run an \textit{equal-cost} experiment in which we extend the coverage baseline to 25 iterations, costing \$220 USD, roughly matching the cost of running \our for 10 iterations (\$211 USD).
%
%
%
We cross-compare their resulting translations using \textit{relative pass rate}.
The two methods complete different numbers of iterations, resulting in different numbers of test cases.
However, even under equal budget, the coverage baseline does not achieve comparable correctness.
Its relative pass rate \toadd{(75/177 = 42.4\%)} remains far below that of \our \toadd{(425/447 = 95.1\%)}.
In terms of failure counts, \our fails only 22 out of \toadd{447} tests from the coverage baseline, while the coverage baseline fails 102 out of \toadd{177} from \our.
This demonstrates that the adversarial setting delivers substantially higher correctness per dollar spent.

\section{Ablation Study on Configurations}
\label{app:ablation-config}
We conduct a second ablation study on the micro-benchmark to identify an effective experimental configuration.
We examine three factors: \textit{seed test cases}, \textit{number of iterations}, and \textit{number of test cases} generated by the discriminator agent per iteration.
Recall that the default configuration of \our uses 15 initial seed tests, runs for 10 iterations, and adds 3 new tests per iteration.
To assess the effect of the number of tests added per iteration, we run two variants of \our that follow the default setting except that they add 1 and 5 test(s) per iteration, respectively, instead of 3.
To analyze the relation between correctness and the number of iterations, we run \our for 20 iterations with other parameters set as default.
We collect the initial translation ($0^{th}$ iteration) and the translated code at the end of $5^{th}$, $10^{th}$, $15^{th}$, and $20^{th}$ iterations.
We compute their \textit{pass rates} on the validation test set. The results summed across programs are shown on the right of Figure~\ref{fig:ablation2}.
%
%
%
%
Compared with adding 1 test per iteration, adding 3 new tests costs 16\% more USD to translate all programs, and improves the final pass rate from 94.7\% to 98.2\%.
Adding 5 new tests per iteration further increases the money cost by 29\% compared with adding 3 tests per iteration, but only improves the pass rate by 1.4\%.
When comparing across different iteration numbers, the validation pass rate is improved by 9\% after the first 10 iterations.
Further extending to 20 iterations yields only a 1.4\% improvement, at the expense of 2.1$\times$ the total financial cost.
At last, we analyze the influence of initial test cases. We run \our starting with only a single seed test, while keeping all other parameters as default values.
A single seed test is sufficient to demonstrate the expected test case format and prevent trivial failures due to formatting mismatches, but it provides less initial coverage than the default setting of 15 seed tests.
%
Using the default configuration (15 initial tests) increases cost by 6\% relative to the single-seed setup, and improves the final pass rate from 94.9\% to 98.2\%.
To balance performance and cost, we adopt the configuration of 15 seed tests, 10 iterations, and 3 new tests per iteration when translating the macro-benchmark.


\section{Macro Experiments with Claude Code and Claude-Sonnet-4}
\label{sec:oldmacro}
\begin{figure*}
\centering
\includegraphics[width=0.99\textwidth]{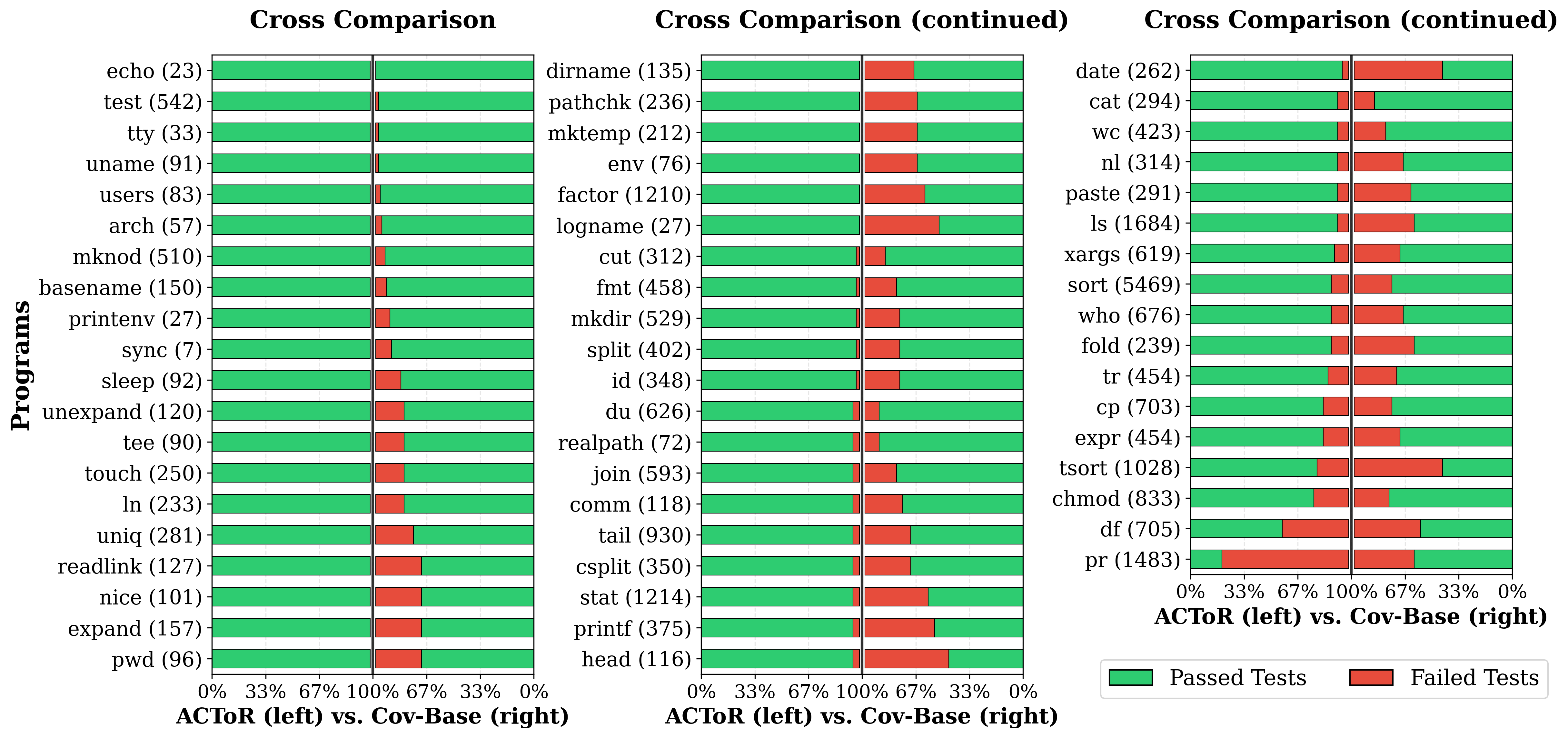}   
\caption{
The relative pass rate when cross-comparing \our and coverage baseline (Cov-Base) on Claude Code with Claude-Sonnet-4 at iteration 10. For each program, the left bar shows evaluating the translation from \our on tests generated in coverage-baseline; the right bar is evaluating the translation from coverage-baseline on \our's tests. The length of each program in LoC is presented next to the program name.
}
\label{fig:macro-res}
\end{figure*}

We also run \our with Claude-Sonnet-4 to translate the macro-benchmark.
In this trial, we use an older implementation of \our that discourages the use of unsafe, but does not enforce full safe Rust in the middle of the translation loop. Instead, full safe Rust is enforced at the end of the translation as a post-processing step, performed via a single agent call that eliminates all \texttt{unsafe} blocks while preserving the functionality of the translated Rust code. The prompt for this post-processing step is shown in Figure~\ref{fig:promptpost}.
We emphasize that the latest implementation of \our used in the main paper (with Claude-Sonnet-4.5) does \emph{not} include this post-processing step: full safety is instead enforced at every iteration of the translation loop, as described in Sec.~\ref{sec:atl}.
All other experimental settings in this trial are identical to those described in the main text.
The results are presented in Figure~\ref{fig:macro-res}.
With Claude-Sonnet-4, \our outperforms the coverage baseline on 55/57 programs in terms of relative pass rate. %
\our achieves full pass on 26 programs using the tests from the coverage baseline.  
Across all 57 programs, \our attains an average relative pass rate of 93.9\%, which is 18.9\% higher than the 75.0\% pass rate of the coverage baseline.
%
\our achieves a 95.3\% pass rate on the union of the test cases from both \our and the coverage baseline.
%
%
This result aligns with the findings on Claude-Sonnet-4.5, demonstrating that \our's performance is stable and generalizes across different model configurations.


\section{Mismatch Detection through Agent-driven Differential Fuzzing}
\label{sec:diffuzz}

\begin{figure}[h]
\centering
\includegraphics[width=\textwidth]{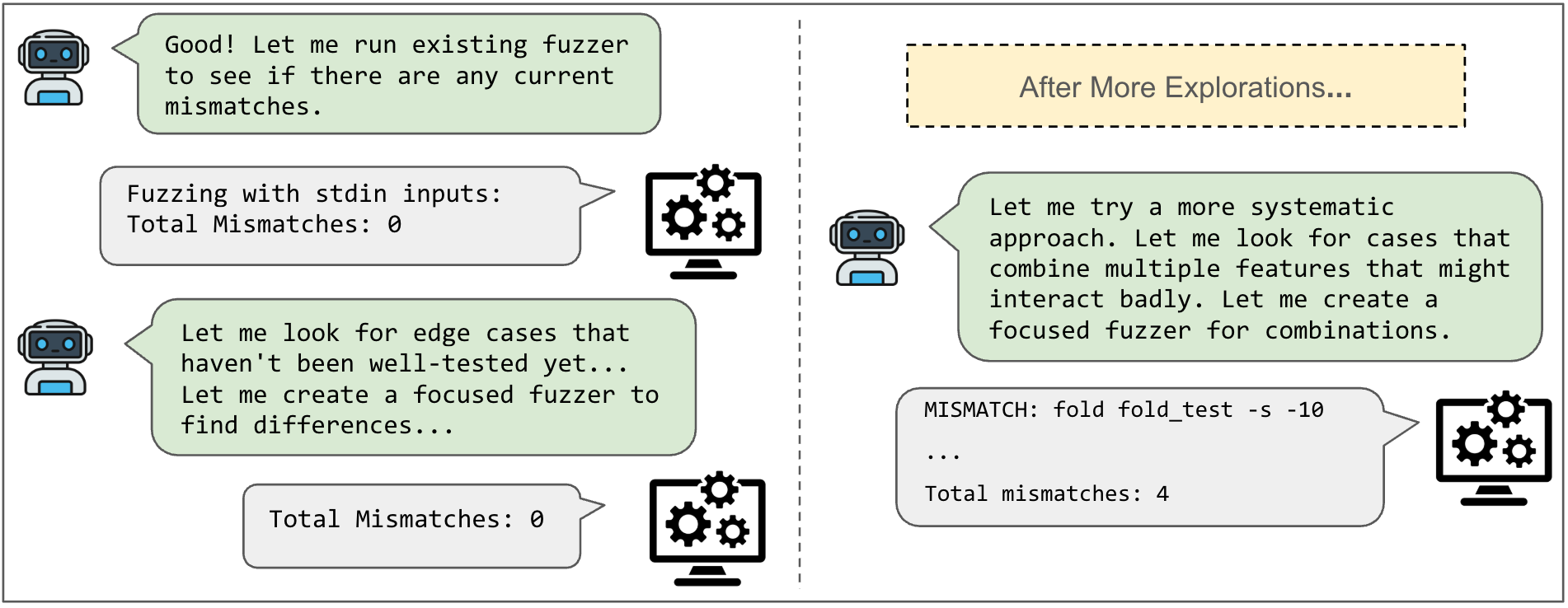}
\caption{One example fuzzer updating during a discrimination procedure. The discriminator agent keeps updating the fuzzer to find mismatches.}
\label{fig:agentfuzz}
\end{figure}

The discrimination procedure aims to detect semantic mismatches between the C code and its Rust translation, thereby guiding the translator agent to improve translation quality.
To effectively uncover functional discrepancies across the two implementations, we incorporate two key design choices.
First, as mentioned in Section~\ref{sec:atl}, we employ an LLM agent as the mismatch detector across C code and Rust code.
Second, rather than relying solely on static source-code inspection or manual testing, we prompt the discriminator to construct and refine a fuzzer with customized input generators that differentially exercise both implementations across a diverse input space.
Recent work on LLM-assisted software testing and fuzzing has shown that LLMs can leverage their understanding of program semantics and specifications to generate focused and structured input cases~\citep{fuzzforall, llm4fuzz1, llm4fuzz2, llm4test1, fuzzsyn2}.
Consistent with these findings, our manual inspection (presented in Appendix~\ref{sec:casestudy}) shows that the LLM agent frequently updates the fuzzer to improve mismatch detection during the discrimination procedure.
Figure~\ref{fig:agentfuzz} presents one representative update that happened in the \code{fold} program.
The discriminator begins by running the existing fuzzer inherited from the previous iteration. After determining that this fuzzer fails to uncover new mismatches, the discriminator controls the fuzzer to probe previously unexplored regions.
The updated fuzzer initially still reports no discrepancies.
Through another refinement of the fuzzer, the discriminator eventually identifies four distinct behaviour mismatches.
These observations suggest that the discriminator needs to continuously and adaptively manipulate the fuzzer to detect mismatches during the adversarial translation process.
Furthermore, as demonstrated in the ablation study in Section~\ref{sec:ablation}, incorporating fuzzer construction in the discrimination procedure improves overall performance by 2\% while reducing total LLM query costs by approximately 12\%, compared to prompting the discriminator to identify mismatches through manual testing alone.
This shows that the tailored fuzzer improves the LLM agent's cost efficiency at detecting mismatches.

\section{Case Study on the Agentic Discriminator}
\label{sec:casestudy}
\fix{
To better understand the role played by the LLM agent in the discrimination procedure, we conduct a case study by analyzing detailed interactions between the agent and the constructed fuzzer using agent execution logs.
We first examine how frequently the LLM agent interacts with the fuzzer, measured by the number of edits on the fuzzer.
Specifically, we identify such interactions using the \texttt{edit/write} tool calls recorded in the logs.
On the 6 programs in the micro benchmark, the agent edits the fuzzer script 126 times over 10 iterations, averaging approximately 21 edits per program.
On the macro benchmark consisting of 57 programs, the fuzzer is updated 312 times in total across 10 iterations, with an average of 5.5 edits per program.
We then conduct a manual inspection of the interaction patterns between the discriminator agent and the fuzzer.
We sample 10 programs (3 from the micro benchmark, 7 from the macro benchmark) and analyze the discriminator agent logs for all 10 iterations.
There are 127 fuzzer edits in total during the adversarial translation process for these 10 programs.
Based on the underlying motivations, we categorize the fuzzer updates into three groups:
(1) \textbf{Redirection.} After inspecting the code or examining existing test cases, the agent determines that the current fuzzer is no longer effective at uncovering new mismatches. It then updates the fuzzer to target different regions of the input space that may expose additional discrepancies. There are 94 cases in this category.
(2) \textbf{Targeted Refinement.} After confirming a semantic mismatch during broad exploration, the discriminator further refines the fuzzer to target the identified pattern and its logically related inputs. 22 case belongs to this category.
(3) \textbf{Unrelated Update.} The remaining 11 edits are cases where the discriminator agent fixes bugs in the fuzzer, separates one update into multiple edits, etc., that are not related to mismatch detection.
Based on these inspections, the fuzzer needs to be constantly updated to discover new discrepancies during the adversarial translation process. A fixed fuzzer with a fixed set of seed tests may not be sufficient to find new mismatches.
}



\end{document}